\documentclass[preprint,showpacs,preprintnumbers,amssymb,aps,nofootinbib]{revtex4}
\usepackage{amsfonts,color,graphicx,epsfig,graphicx,amsmath,multirow,slashed}

\begin{document}

\title{Doubly-charmed baryon production in $Z$ boson decay}
\author{Xuan Luo}
\author{Ying-Zhao Jiang}
\author{Gui-Yuan Zhang}
\author{Zhan Sun}
\email{zhansun@cqu.edu.cn}
\affiliation{
Department of Physics, Guizhou Minzu University, Guiyang 550025, People's Republic of China.
}

\date{\today}

\begin{abstract}
In this paper, we carry out a detailed study of doubly-charmed baryon production in $Z$ boson decay, on the basis of the nonrelativistic QCD factorization. With the inclusion of the di-quark states $(cc)[^3S_1]_{\bar{\textbf{3}}}$ and $(cc)[^1S_0]_{{\textbf{6}}}$, the branching ratio of $\mathcal{B}_{Z \to \Xi_{cc}+X}$ is predicted to be of the $10^{-5}$ order, indicating its experimental measurability. By comparing to the $\Lambda^{+}_{c}$ yield in $Z$ decay, we predict $\mathcal{R}_{\Xi_{cc}^{+}}(=\frac{\Gamma(Z \to \Xi^{+}_{cc}) \times \mathcal{B}(\Xi^{+}_{cc} \to \Lambda_c^{+}K^{-}\pi^{+})}{\Gamma(Z \to \Lambda_c^{+})})=(0.85^{+0.10}_{-0.07}) \times 10^{-4}$ and $\mathcal{R}_{\Xi_{cc}^{++}}(=\frac{\Gamma(Z \to \Xi^{++}_{cc}) \times \mathcal{B}(\Xi^{++}_{cc} \to \Lambda_c^{+}K^{-}\pi^{+}\pi^{+})}{\Gamma(Z \to \Lambda_c^{+})})=(1.70^{+0.20}_{-0.14}) \times 10^{-4}$, which are at clear variance with the SELEX measurements but comparable with the values given by the LHCb and Belle collaborations.

\end{abstract}

\maketitle

\section{Introduction}

The doubly-charmed baryon (labeled as $\Xi_{cc}$), which is assumed to contain two $c$ quarks and a light quark $q$ ($q=u,d,s$) based on the quark model \cite{quark model 1,quark model 2,quark model 3,quark model 4,quark model 5}, can provide unique test for the quantum chromodynamics (QCD). The past decades have seen the rapid developments of the $\Xi_{cc}$ related studies, including both experimental and theoretical aspects.

By reconstructing $\Xi_{cc}^{+}$ ($ccu$) via its decay into $\Lambda_{c}^{+}K^{-}\pi^{+}$, the SELEX collaboration reported a large production rate of $\Xi_{cc}^{+}$ (i.e. $\mathcal{R}_{\Xi_{cc}^{+}}=\frac{\sigma(\Xi^{+}_{cc}) \times \mathcal{B}(\Xi^{+}_{cc} \to \Lambda_c^{+}K^{-}\pi^{+})}{\sigma(\Lambda_c^{+})}=9\%$) \cite{SELEX:2002wqn}, which, however, was even not confirmed by the FOCUS collaboration \cite{Ratti:2003ez} that is at the same collider of SELEX. In 2013, the LHCb Collaboration performed its first search for $\Xi_{cc}^{+}$, reporting the upper limit of $\mathcal{R}_{\Xi_{cc}^{+}}$ is from $1.5 \times 10^{-2}$ (corresponding to the $\Xi_{cc}^{+}$ lifetime of 100 fs) to $3.9 \times 10^{-4}$ (400 fs) \cite{LHCb:2013hvt}, which is recently updated to be from $6.5 \times 10^{-3}$ (40 fs) to $9 \times 10^{-4}$ (160 fs) for $\sqrt{s}=8$ TeV, and from $4.5 \times 10^{-4}$ (40 fs) to $1.2 \times 10^{-4}$ (160 fs) for $\sqrt{s}=13$ TeV \cite{LHCb:2019gqy}. The LHCb data appear to be at clear variance with the SELEX-measured $\mathcal{R}_{\Xi_{cc}^{+}}$. Comparing to $\Xi_{cc}^{+}$, $\Xi_{cc}^{++}$ ($uud$) has a much longer lifetime and is then much easier to be tagged from the LHC background \cite{Wu:2019gta}. In 2017, the LHCb collaboration firstly detected the decay channel $\Xi_{cc}^{++} \to \Lambda_{c}^{+} K^{-} \pi^{+}\pi^{+}$ with $\Lambda_{c}^{+} \to p K^{-} \pi^{+}$ \cite{LHCb:2017iph}, and then observed the decay channels of $\Xi_{cc}^{++} \to \Xi_{cc}^{+} \pi^{+}$ \cite{LHCb:2018pcs} and $\Xi_{cc}^{++} \to \Xi_{cc}^{+} \pi^{+} \pi^{+}$ \cite{LHCb:2018zpl}. In 2019, LHCb achieved the first measurement of the $\Xi_{cc}^{++}$ production in proton-proton collision \cite{LHCb:2019qed}, reporting $\mathcal{R}_{\Xi_{cc}^{++}}(=\frac{\sigma(\Xi^{++}_{cc}) \times \mathcal{B}(\Xi^{++}_{cc} \to \Lambda_c^{+}K^{-}\pi^{+}\pi^{+})}{\sigma(\Lambda_c^{+})})=(2.22 \pm 0.27 \pm 0.29) \times 10^{-4}$. Besides the hadroproduction, the $\Xi_{cc}$ production in $e^{+}e^{-}$ annihilation has also been measured \cite{BaBar:2006bab,Belle:2006edu,Belle:2013htj}. The upper limits of $\sigma(e^{+}e^{-} \to \Xi_{cc}^{+(++)}+X)$ given by the Belle collaboration are comparable with the theoretical predictions; the values of $\mathcal{R}_{\Xi_{cc}^{+(++)}}$ are measured to be $\sim 10^{-4}$.

In addition to the direct productions \cite{direct 1,direct 2,direct 3,direct 4,direct 5,direct 6,direct 7,direct 8,direct 9,direct 10,direct 11,direct 12,direct 13,direct 14,direct 15,direct 16,direct 17,direct 18,direct 19,direct 20,direct 21,direct 22,direct 23,direct 24,direct 25,direct 26,direct 27,direct 28,direct 29,direct 30,direct 31,direct 32,direct 33,direct 34,direct 35,direct 36,direct 37}, such as the hadro-, photo-, and electroproductions, the indirect $\Xi_{cc}$ yield in decays \cite{indirect 1,indirect 2,indirect 3,indirect 4}, which is indeed very complementary to the direct case, is also of great interest to study the doubly-charmed baryon. For example, Niu. $et~al.$ pointed out about $10^{3}$ $\Xi_{cc}$ events can be accumulated through higgs decay in one running year at the proposed HL-LHC; Li. $et~al.$ indicated the branching ratio of $\mathcal{B}_{\Upsilon(1S) \to \Xi_{cc}+X}$ can be significant and can be well measured as the $\Upsilon(1S)$ decay to $J/\psi+c+\bar{c}$+X. In addition to these decay processes, the $Z$ boson decay could also provide a uniquely good chance for the $\Xi_{cc}$-related study. At the LHC, $\sim10^{9}$ $Z$ events are expected to be generated per year \cite{Liao:2015vqa}, which would be largely increased by the HE(L)-LHC upgrade program. The proposed future $e^+e^-$ collider, CEPC \cite{CEPC}, equipped with $``\textrm{clean}"$ background and enormous $Z$ production events ($\sim10^{12}$/year), would also be beneficial to hunt $\Xi_{cc}$ yield through $Z$ decay. Thus, it appears promising to measure $Z$ decaying into inclusive $\Xi_{cc}$. Moreover, heavy-quarkonium production in $Z$ decay (such as $Z \to J/\psi+X$), which is analogue to our concerned $\Xi_{cc}$ case, has triggered increasing attentions and has accumulated abundant experimental data \cite{Zyla:2020zbs}. Taken together, we, in this paper, would perform a detailed study of $Z \to \Xi_{cc}+X$, presenting the estimations of $\mathcal{R}_{\Xi_{cc}^{+(++)}}$ in the course of $Z$ decay.

The rest of the paper is organized as follows: In Sec. II, we give a description on the calculation formalism. In Sec. III, the phenomenological results and discussions are presented. Section IV is reserved as a summary.

\section{Calculation Formalism}

\subsection{General Formalism}
The production of the doubly-charmed baryon is often assumed to be factorized into two steps \cite{direct 1,direct 11,direct 27}. The first procedure is to produce a $c$-quark pair $(cc)[n]$ by the perturbative calculable hard processes, with subsequent nonperturbative transition into a bounding diquark $\langle cc \rangle [n]$ that can be described by a matrix element, $h_{[n]}$. The next step is the hadronization of $\langle cc \rangle [n]$ into a physical colorless baryon $\Xi_{ccq}$ ($q=u,d,s$) by grabbing a light quark with possible soft gluons from the hadron; during the hadronization, the $``\textrm{direct~evolution~mechanism}"$ assumes the total evolving probability to be $100\%$, among which the fragmentation probabilities into $\Xi_{ccu}$, $\Xi_{ccd}$, and $\Xi_{ccs}$ account for $43\%$, $43\%$, and $14\%$, respectively \cite{Wu:2019gta,direct 30,Sjostrand:2006za}.

Within the nonrelativistic QCD (NRQCD) framework \cite{NRQCD}, the differential decay width of $Z \to \Xi_{cc}+X$ can be expressed as
\begin{eqnarray}
d\Gamma=d\hat{\Gamma}_{Z \to (cc)[n]+X}\langle \mathcal O ^{\Xi_{cc}}(n)\rangle,\label{fac exp}
\end{eqnarray}
where $\hat{\Gamma}_{Z \to (cc)[n]+X}$ is the perturbative calculable short distance coefficients (SDCs), representing the production of a configuration of the $(cc)[n]$ intermediate state. At the leading order of $v_c$ (the relative velocity of the two constituent $c$ quarks in the diquark),\footnote{The contributions of the $P$-wave processes would at least be $v_c^2$ suppressed to that of the $S$-wave processes. For example, the $P$-wave contributions just account for about $3\%-5\%$ of the total cross sections of the hadroproduced $\Xi_{cc}$ \cite{Berezhnoy:2020aox}.} $n$ takes on either $[^3S_1]_{\bar{\textbf{3}}}$ or $[^1S_0]_{{\textbf{6}}}$, due to the symmetry of identical particles in the diquark state. The subscripts $\bar{\textbf{3}}({\textbf{6}})$ (as will be depicted in later section) denote the color state of the diquark.  The universal long distance matrix element $\langle\mathcal O^{\Xi_{cc}}(n)\rangle$ stands for the transition probability of the $c$-quark pair into the diquark multiplied by the subsequent fragmentation probability into $\Xi_{cc}$, i.e., $h_{([^3S_1]_{\bar{\textbf{3}}},[^1S_0]_{{\textbf{6}}})}\times 100\%$ based on the ``direct evolution mechanism".

\subsection{Amplitudes}

\begin{figure*}
\includegraphics[width=0.95\textwidth]{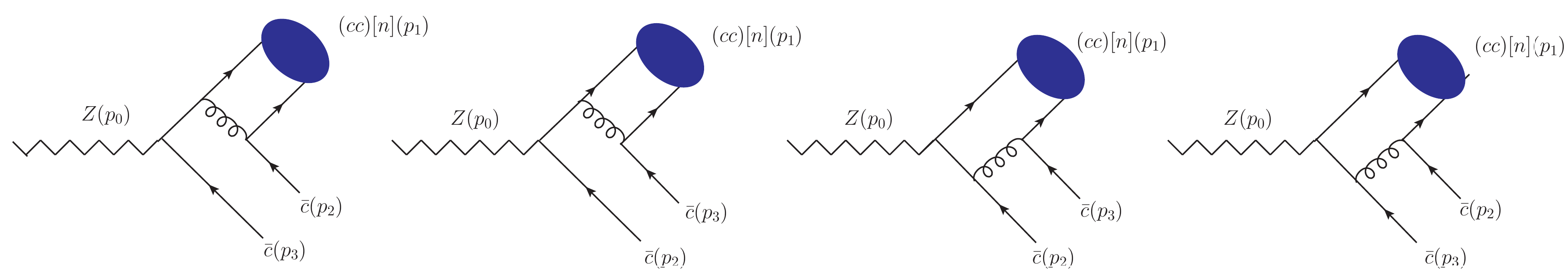}
\caption{\label{fig:Feynman}
Typical Feynman diagrams of $Z \to (cc)[n]+\bar{c}+\bar{c}$.}
\end{figure*}

The SDCs in Eq. (\ref{fac exp}) can be written as
\begin{eqnarray}
\hat{\Gamma}_{Z \to (cc)[n]+X}=|\mathcal{M}|^{2}d\Phi_3,
\end{eqnarray}
where $|\mathcal{M}|^2$ and $d\Phi_3$ are the squared matrix element and the standard 3-body phase space, respectively. For $Z \to (cc)[n]+\bar{c}+\bar{c}$, there are in total 8 Feynman diagrams, half of which are shown in Fig. 1. The other 4 ones can be obtained by exchanging the two identical $c$-quark lines inside the diquark. Because we have set $v_c=0$, the two portions of 4 diagrams contribute identically; in this case, we only need to calculate the 4 diagrams in Figure \ref{fig:Feynman}, and multiply a factor of $2^2$. Simultaneously we should introduce an additional factor of $1/(2!2!)$ to deal with the identities of the two constituent $c$ quarks inside the diquark and the two final-state $\bar{c}$ quarks.

According Fig. \ref{fig:Feynman}, one can obtain
\begin{eqnarray}
\mathcal{M}_1&=&- \kappa \frac{{ \bar u\left( {{p_{12}}} \right)\left( {-i{\gamma ^{\nu}}} \right)v\left( {{p_2}} \right) \bar u\left( {{p_{11}}} \right)\left( { - i{\gamma ^{\nu}}} \right)\left( {{m_{c}}+{p\!\!\!\slash_{1}}+{p\!\!\!\slash_{2}}} \right) \epsilon\!\!\!\slash(p_0)(c+\gamma^{5}) v\left( {{p_3}} \right)}}{{{{\left( {{p_{12}} + {p_2}} \right)}^2}\left[ {{{\left( {{p_1} + {p_2}} \right)}^2} - m_{c}^2} \right]}},
\nonumber \\
\mathcal{M}_2&=& - \kappa \frac{{ \bar u\left( {{p_{12}}} \right)\left( { - i{\gamma ^{\nu}}} \right)\left( {{m_c}+{p\!\!\!\slash_1}+{p\!\!\!\slash_3}} \right) \epsilon\!\!\!\slash(p_0)(c+\gamma^{5}) v\left( {{p_2}} \right) \bar u\left( {{p_{11}}} \right)\left( { - i{\gamma ^{\nu}}} \right)v\left( {{p_3}} \right)}}{{{{\left( {{p_{11}} + {p_3}} \right)}^2}\left[ {{{\left( {{p_1}+{p_3}} \right)}^2} - m_c^2} \right]}},
\nonumber \\
\mathcal{M}_3&=& - \kappa \frac{{ \bar u\left( {{p_{12}}} \right) \epsilon\!\!\!\slash(p_0)(c+\gamma^{5}) \left( {{m_c} - {p\!\!\!\slash_{11}} - {p\!\!\!\slash_2} - {p\!\!\!\slash_3}} \right)\left( { - i{\gamma ^{\nu}}} \right)v\left( {{p_2}} \right) \bar u\left( {{p_{11}}} \right)\left( { - i{\gamma ^{\nu}}} \right) v\left( {{p_3}} \right)}}{{{{\left( {{p_{11}} + {p_3}} \right)}^2}\left[ {{{\left( {{p_{11}} + {p_2} + {p_3}} \right)}^2} - m_c^2} \right]}},
\nonumber \\
\mathcal{M}_4&=& - \kappa \frac{{\bar u\left( {{p_{12}}} \right)\left( { - i{\gamma ^{\nu}}} \right)v\left( {{p_2}} \right) \bar u\left( {{p_{11}}} \right) \epsilon\!\!\!\slash(p_0)(c+\gamma^{5}) \left( {{m_{c}} - {p\!\!\!\slash_{12}} - {p\!\!\!\slash_2} - {p\!\!\!\slash_3}} \right)\left( { - i{\gamma ^{\nu}}} \right)v\left( {{p_3}} \right)}}{{{{\left( {{p_{12}} + {p_2}} \right)}^2}\left[ {{{\left( {{p_{12}} + {p_2} + {p_3}} \right)}^2} - m_{c}^2} \right]}}, \nonumber \\ \label{ori amp}
\end{eqnarray}
where $\kappa=-\mathcal{C}\frac{{i eg_s}^2}{4\cos\theta_{\textrm{w}} \sin\theta_{\textrm{w}}}$ with $\mathcal{C}$ denoting the color factor. $\epsilon(p_{0})$ is the polarization vector of the initial $Z$ boson. The coefficient $c$ reads
\begin{align}
c =\frac{8}{3}\sin^2\theta_{\textrm{w}}-1.
\end{align}

The momenta of the constituent quarks in the diquark follow as
\begin{eqnarray}
p_{11}=\frac{m_c}{M_{cc}}p_1+q~~\textrm{and}~~p_{12}=\frac{m_c}{M_{cc}}p_1-q,
\end{eqnarray}
where $m_{c}=M_{cc}/2$ is implicitly adopted to ensure the gauge invariance of the hard scattering amplitude; $q( \simeq 0)$ is the relative momentum between the two constituent $c$ quarks inside the diquark.

By inserting the charge conjugate matrix $C=-i \gamma^{2}\gamma^{0}$ that satisfies the following equations \cite{direct 27},
\begin{eqnarray}
&&CC^{-1}=1,~~~v^{T}(p)C=-\bar{u}(p),~~~C^{-}\bar{u}\left( p \right)^{T}
=v\left(p \right), \nonumber \\
&&C^{-} (\gamma^{\mu})^{T} C=-\gamma^{\mu},~~~C^{-} (\gamma^{\mu}\gamma^{5})^{T} C=\gamma^{\mu}\gamma^{5}, \nonumber \\
&&C^{-}s^{T}_f(k,m)C=s_f(-k,m),
\end{eqnarray}
the amplitudes in Eq. (\ref{ori amp}) can be rewritten as
\begin{eqnarray}\label{eq:22}
\mathcal{M}_1&=&- \kappa \frac{{ \bar u\left( {{p_{2}}} \right)\left( {-i{\gamma ^{\nu}}} \right) \Pi^{[n]}_{p_{1}} \left( { - i{\gamma ^{\nu}}} \right)\left( {{m_{c}}+{p\!\!\!\slash_1}+{p\!\!\!\slash_2}} \right) \epsilon\!\!\!\slash(p_0)(c+\gamma^{5}) v\left( {{p_3}} \right)}}{{{{\left( {{p_{12}} + {p_2}} \right)}^2}\left[ {{{\left( {{p_1} + {p_2}} \right)}^2} - m_{c}^2} \right]}},
\nonumber \\
\mathcal{M}_2&=& - \kappa \frac{{\bar u\left( {{p_{2}}} \right)\left( { - i{\gamma ^{\nu}}} \right)\left( {{m_c}-{p\!\!\!\slash_1}-{p\!\!\!\slash_3}} \right) \epsilon\!\!\!\slash(p_0)(c-\gamma^{5}) \Pi^{[n]}_{p_{1}} \left( { - i{\gamma ^{\nu}}} \right)v\left( {{p_3}} \right)}}{{{{\left( {{p_{11}} + {p_3}} \right)}^2}\left[ {{{\left( {{p_1}+{p_3}} \right)}^2} - m_c^2} \right]}},
\nonumber \\
\mathcal{M}_3&=& - \kappa \frac{{\bar u\left( {{p_{2}}} \right)\epsilon\!\!\!\slash(p_0)(c-\gamma^{5})\left( {{m_c} + {p\!\!\!\slash_{11}} + {p\!\!\!\slash_2} + {p\!\!\!\slash_3}} \right)\left( { - i{\gamma ^{\nu}}} \right) \Pi^{[n]}_{p_{1}} \left( { - i{\gamma ^{\nu}}} \right) v\left( {{p_3}} \right)}}{{{{\left( {{p_{11}} + {p_3}} \right)}^2}\left[ {{{\left( {{p_{11}} + {p_2} + {p_3}} \right)}^2} - m_c^2} \right]}},
\nonumber \\
\mathcal{M}_4&=& - \kappa\frac{{ \bar u\left( {{p_{2}}} \right)\left( { - i{\gamma ^{\nu}}} \right) \Pi^{[n]}_{p_{1}} \epsilon\!\!\!\slash(p_0)(c+\gamma^{5}) \left( {{m_{c}} - {p\!\!\!\slash_{12}} - {p\!\!\!\slash_2} - {p\!\!\!\slash_3}} \right)\left( { - i{\gamma ^{\nu}}} \right)v\left( {{p_3}} \right)}}{{{{\left( {{p_{12}} + {p_2}} \right)}^2}\left[ {{{\left( {{p_{12}} + {p_2} + {p_3}} \right)}^2} - m_{c}^2} \right]}},
\end{eqnarray}
where $\Pi^{[n]}_{q}$ denotes the spin projector operators \cite{Petrelli:1997ge},
\begin{eqnarray}
\Pi^{[^1S_0]}_{q}&=&\frac{1}{\sqrt{8m_c^3}} \left(\frac{q\!\!\!\slash}{2}-m_c\right) \gamma^{5} \left(\frac{q\!\!\!\slash}{2}+m_c\right), \nonumber \\
\Pi^{[^3S_1]}_{q}&=&\frac{1}{\sqrt{8m_c^3}} \left(\frac{q\!\!\!\slash}{2}-m_c\right) \epsilon\!\!\!\slash_q \left(\frac{q\!\!\!\slash}{2}+m_c\right).
\end{eqnarray}

\subsection{Color Factor}

According to Fig. \ref{fig:Feynman}, the color factor $\mathcal{C}$ can be expressed as
\begin{eqnarray}
{\mathcal{C}} =T^{a}_{im}T^{a}_{jn},
\end{eqnarray}
where $a=1 \cdots 8$ is the color indices of the incoming gluon; $i,j=1,2,3$ and $m,n=1,2,3$ denote the color indices of the two constituent $c$ quarks in the diquark and that of the two final-state $\bar{c}$ quarks, respectively. By the fact that $3 \otimes 3=\bar{\textbf{3}} \oplus \textbf{6}$ in $\textrm{SU}_{\textrm{c}}(3)$ group, the diquark can be either in anti-triplet $\bar{\textbf{3}}$ or in sextuplet $\textbf{6}$ color state; in this case, we introduce the function $\frac{G_{ijk}}{N}$ to describe the diquark color, $k=3$ and $N=\sqrt{2}$ being the color indices of the diquark and the normalized factor, respectively. $G_{ijk}$ is identical to the antisymmetric $\varepsilon_{ijk}$ ($\bar{\textbf{3}}$) or the symmetric $f_{ijk}$ ($\textbf{6}$),
which satisfies the following equations
\begin{eqnarray}
\varepsilon_{ijk}\varepsilon_{i^{'}j^{'}k}&=&\delta_{i i^{'}}\delta_{j j^{'}}-\delta_{j i^{'}}\delta_{i j^{'}}, \nonumber \\
f_{ijk}f_{i^{'}j^{'}k}&=&\delta_{i i^{'}}\delta_{j j^{'}}+\delta_{j i^{'}}\delta_{i j^{'}}.
\label{color}
\end{eqnarray}

\section{Phenomenological results}

The input parameters in our calculations are set as
\begin{eqnarray}
m_Z&=&91.1876~\textrm{GeV},~~~m_{c}=1.8 \pm 0.05~\textrm{GeV},\nonumber \\
\sin^{2}\theta_\textrm{w}&=&0.23116,~~~\alpha=1/137.
\end{eqnarray}
According to the velocity scaling rule of NRQCD, we adopt the usual assumption that the matrix elements $h_{[^3S_1]_{\bar{\textbf{3}}}}$ and $h_{[^1S_0]_{{\textbf{6}}}}$ have the equal values \cite{direct 1,direct 11,direct 27}, which are given by the wave function at the origin \cite{direct 5,direct 27}
\begin{eqnarray}
h_{[^3S_1]_{\bar{\textbf{3}}}}=h_{[^1S_0]_{{\textbf{6}}}}=\big|\Psi_{cc}(0)\big|^2=0.039~\textrm{GeV}^{3}.
\end{eqnarray}

\begin{table*}[htb]
\caption{Decay widths (in units of $\textrm{KeV}$) of $Z \to \Xi_{cc}+X$.}
\label{decay width}
\begin{tabular}{ccccccccc}
\hline\hline
$\mu_r$ & $m_c$ (GeV) & $~~~\Gamma_{[^3S_1]_{\bar{\textbf{3}}}}~~~$ & $~~~\Gamma_{[^1S_0]_{{\textbf{6}}}}~~~$ & $~~~\Gamma_{\textrm{Total}}~~~$ & $~~~\mathcal{B}(\times 10 ^{-5})~~~$\\ \hline
$~$ & $1.75$ & $16.36$ & $58.87$ & $75.23$ & $3.015$\\
$2m_c$ & $1.80$ & $14.59$ & $52.51$ & $67.10$ & $2.689$\\
$~$ & $1.85$ & $13.33$ & $48.01$ & $61.34$ & $2.458$\\ \hline
$~$ & $1.75$ & $4.79$ & $17.21$ & $22.00$ & $0.882$\\
$m_Z/2$ & $1.80$ & $4.40$ & $15.82$ & $20.22$ & $0.810$\\
$~$ & $1.85$ & $4.05$ & $14.57$ & $18.62$ & $0.746$\\ \hline\hline
\end{tabular}
\end{table*}

We summarize the predicted decay widths of $Z \to \Xi_{cc}+X$ in Tab. \ref{decay width}. Inspecting the data, one can find the branching ratio of $Z \to \Xi_{cc}+X$ amounts to $\sim 10^{-5}$, which is comparable with the color-singlet predictions of $\mathcal{B}_{Z \to J/\psi+X}$ \cite{Barger:1989cq,Braaten:1993mp}. In the predictions of the total decay width, the state of $(cc)[^3S_1]_{\bar{\textbf{3}}}$ plays the leading role, more than three times bigger in magnitude than that of $(cc)[^1S_0]_{{\textbf{6}}}$. Varying $m_c$ around the default value of $1.8$ GeV by $\pm 0.05$ GeV would arouse a $10\%$ variation of the decay width.

Based on the aforementioned capability of hunting $Z$ boson at LHC and CEPC, about $10^{4}$ (LHC) and $10^{7}$ (CEPC) $\Xi_{cc}$ production events arisen from $Z$ decay would be collected in one running year at the two colliders. By further considering the decay chains of $\Xi_{cc}^{+} \to \Lambda_{c}^{+}K^{-}\pi^{+}$ ($\simeq 5\%$ \cite{LHCb:2013hvt}) or $\Xi_{cc}^{++} \to \Lambda_{c}^{+}K^{-}\pi^{+}\pi^{+}$ ($\simeq 10\%$ \cite{LHCb:2017iph,Yu:2017zst}) with the cascade decay $\Lambda_{c}^{+} \to pK^{-}\pi^{+}$($\simeq 5\%$ \cite{LHCb:2013hvt}), we would accumulate about $10$ and $10^{4}$ reconstructed $\Xi^{+(++)}_{cc}$ events at LHC and CEPC, respectively. Note that, the proposed HL(E)-LHC upgrade program may largely increase the $\Xi_{cc}$ yield events.

We are now in a position to estimate the ratio of the production rate of $\Xi_{cc}^{+(++)}$ in $Z$ decay to that of $\Lambda_{c}^{+}$. According to $\mathcal{B}_{Z \to \Lambda_{c}^{+}}=\mathcal{B}_{Z \to c \bar{c}} \times f_{c \to \Lambda_{c}^{+}}=0.12 \times 0.057=6.84 \times 10^{-3}$, \cite{Zyla:2020zbs,Gladilin:2014tba} we have
\begin{eqnarray}
\mathcal{R}_{\Xi_{cc}^{+}}&=&\frac{\Gamma(Z \to \Xi^{+}_{cc}) \times \mathcal{B}(\Xi^{+}_{cc} \to \Lambda_c^{+}K^{-}\pi^{+})}{\Gamma(Z \to \Lambda_c^{+})}=0.85^{+0.10}_{-0.07} \times 10^{-4}, \nonumber \\
\mathcal{R}_{\Xi_{cc}^{++}}&=&\frac{\Gamma(Z \to \Xi^{++}_{cc}) \times \mathcal{B}(\Xi_{cc}^{++} \to \Lambda_{c}^{+}K^{-}\pi^{+}\pi^{+})}{\Gamma(Z \to \Lambda_c^{+})}=1.70^{+0.20}_{-0.14} \times 10^{-4},
\end{eqnarray}
where $\mu_r=2m_c$ and the uncertainties are caused by $m_c=1.8 \pm 0.05$ GeV. From the ratios one can perceive the predicted $\mathcal{R}_{\Xi_{cc}^{+(++)}}$ in $Z$ decay is comparable with the measurements of LHCb (13 TeV) \cite{LHCb:2019gqy,LHCb:2019qed} and Belle collaborations \cite{BaBar:2006bab}, while still significantly below the SELEX-reported $\mathcal{R}_{\Xi_{cc}^{+}}$ \cite{SELEX:2002wqn}.

\begin{figure*}
\includegraphics[width=0.49\textwidth]{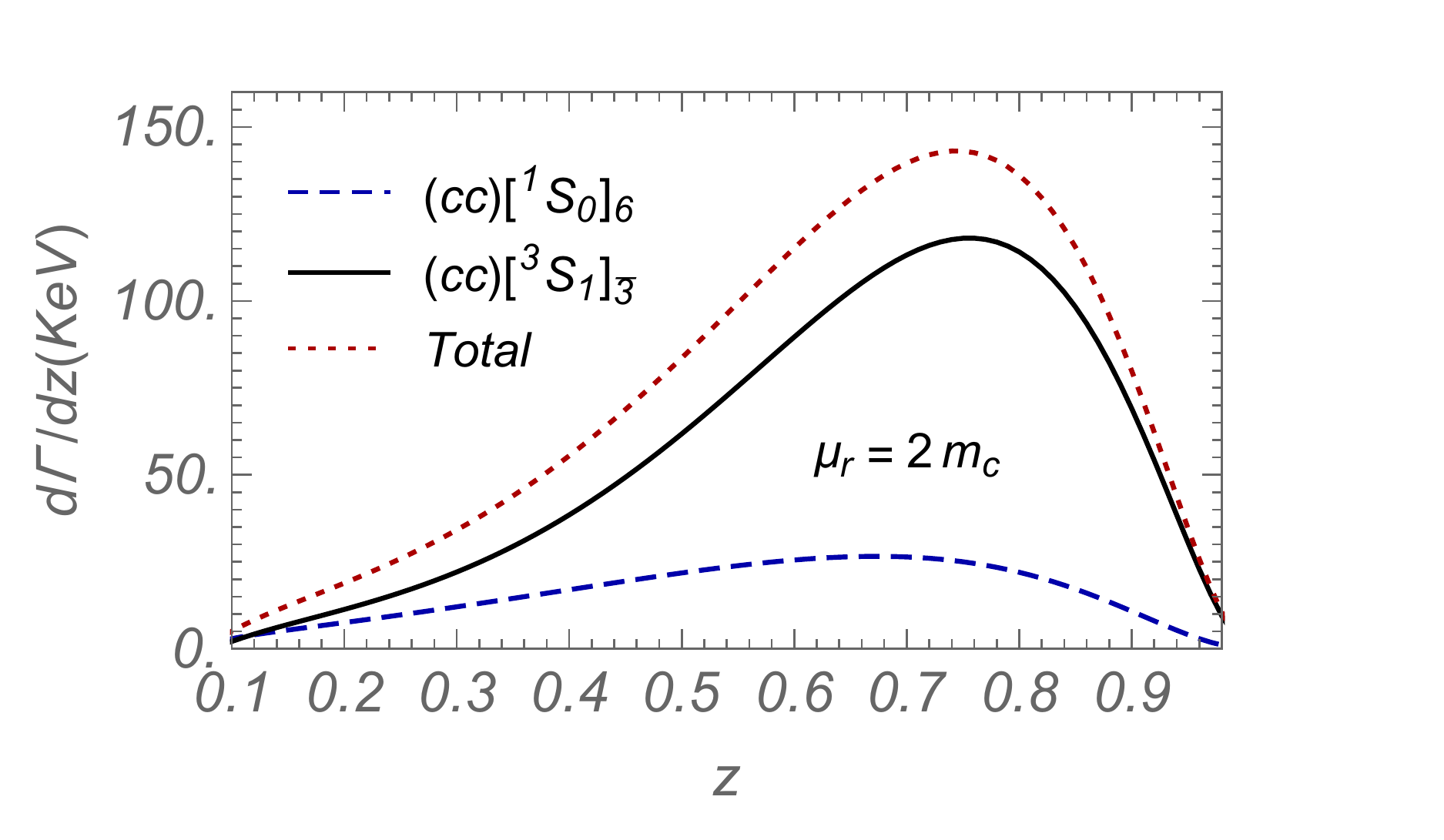}
\includegraphics[width=0.49\textwidth]{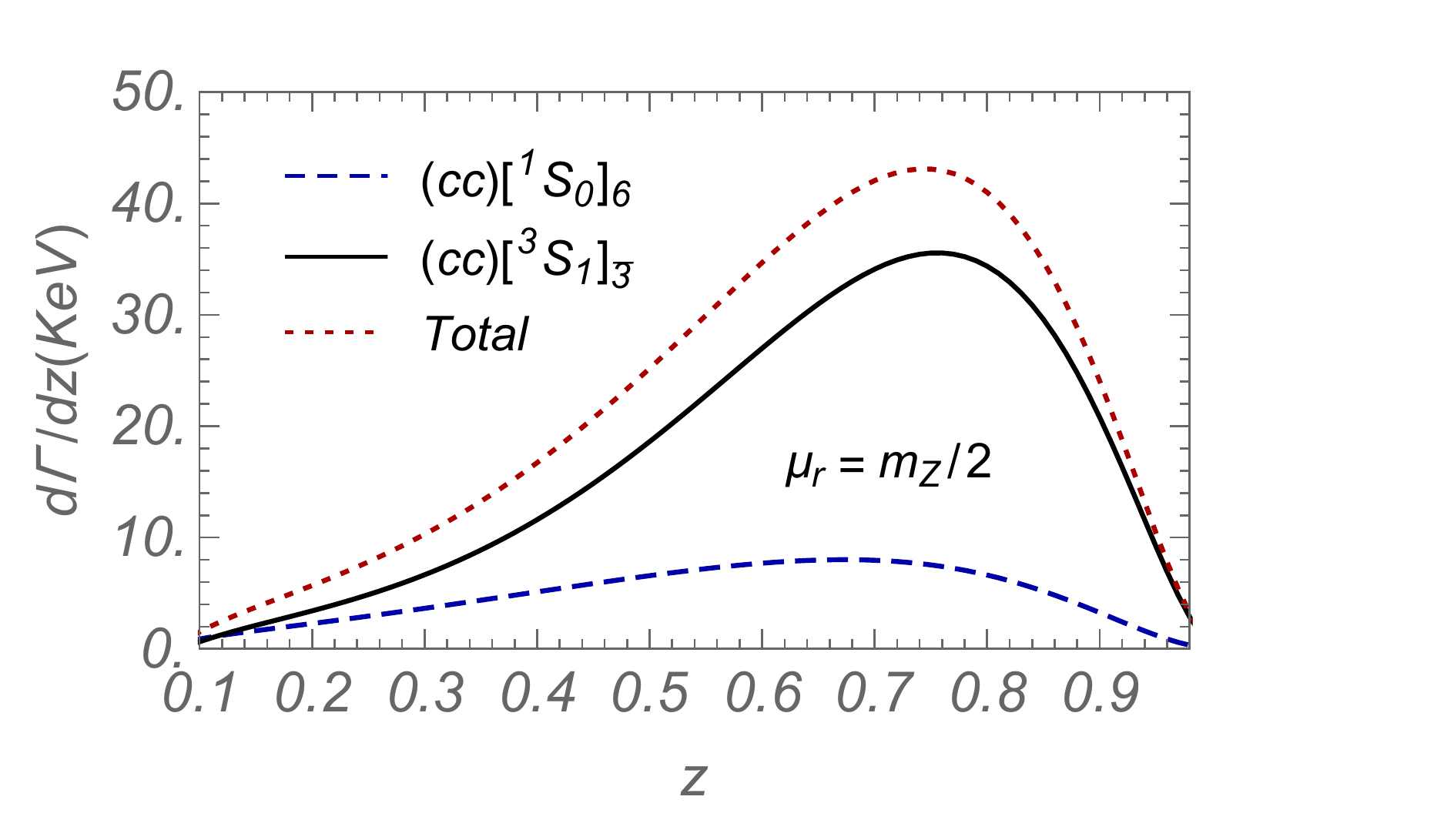}
\caption{\label{fig:z dis}
$\Xi_{cc}$ energy distributions in $Z \to \Xi_{cc}+X$ with $z$ defined as $\frac{2E_{\Xi_{cc}}}{m_Z}$.}
\end{figure*}

At last, we give the predictions of the $\Xi_{cc}$ energy distributions in Fig. \ref{fig:z dis}. The peak of $\frac{d\Gamma}{dz}\big|_{(cc)[^3S_1]_{\bar{\textbf{3}}}}$ is around $z=0.75$ and $\frac{d\Gamma}{dz}\big|_{(cc)[^1S_0]_{\textbf{6}}}$ peaks near $z=0.7$. That the peak of $\Xi_{cc}$ energy distribution in $Z \to \Xi_{cc}+X$ lies in the large $z$ region can primarily be attributed to the dominance of the $c$-quark fragmentation mechanism.

\section{Summary}

In this manuscript, we apply the NRQCD factorization to study the $Z$ boson decaying into inclusive doubly-charmed baryon. By including the contributions of the di-quark states $(cc)[^3S_1]_{\bar{\textbf{3}}}$ and $(cc)[^1S_0]_{{\textbf{6}}}$, the branching ratio of $\mathcal{B}_{Z \to \Xi_{cc}+X}$ is predicted to be $\sim 10^{-5}$, following which as high as $10^{4(7)}$ $\Xi_{cc}$ events from $Z$ decay are expected to be accumulated at LHC (CEPC) per year. By comparing to the measurements on $\mathcal{B}_{Z \to \Lambda^{+}_{c}+X}$, we predict $\mathcal{R}_{\Xi_{cc}^{+}}(=\frac{\Gamma(Z \to \Xi^{+}_{cc}) \times \mathcal{B}(\Xi^{+}_{cc} \to \Lambda_c^{+}K^{-}\pi^{+})}{\Gamma(Z \to \Lambda_c^{+})})=(0.85^{+0.10}_{-0.07}) \times 10^{-4}$ and $\mathcal{R}_{\Xi_{cc}^{++}}(=\frac{\Gamma(Z \to \Xi^{++}_{cc}) \times \mathcal{B}(\Xi^{++}_{cc} \to \Lambda_c^{+}K^{-}\pi^{+}\pi^{+})}{\Gamma(Z \to \Lambda_c^{+})})=(1.70^{+0.20}_{-0.14}) \times 10^{-4}$, where the uncertainties are estimated using alternative choices of the $c$-quark mass. Our prediction of $\mathcal{R}_{\Xi_{cc}^{+(++)}}$ in the course of $Z$ decay is still inconsistent with the SELEX data but compatible with the LHCb- and Belle-measured values.

\section{Acknowledgments}
\noindent{\bf Acknowledgments}:
This work is supported in part by the Natural Science Foundation of China under the Grant No. 12065006, and by the Project of GuiZhou Provincial Department of Science and Technology under Grant No. QKHJC[2020]1Y035. and No. QKHJC[2019]1167.\\

\end{document}